\documentclass[
 reprint,
superscriptaddress,
nofootinbib,
 amsmath,amssymb,
 aps,
 pra,
]{revtex4-2} 
\usepackage{graphicx}
\usepackage{dcolumn}
\usepackage{bm}
\usepackage{comment}
\usepackage{lineno}
\usepackage{siunitx}
\usepackage{import}
\usepackage{amssymb}
\usepackage{epsfig}
\usepackage{hyperref}

\hypersetup{
colorlinks=true,
linkcolor=blue,
citecolor=blue,
urlcolor=blue,
}

\DeclareSIUnit \clight  {\textit{c}}
\RequirePackage[normalem]{ulem} 
\RequirePackage{color}\definecolor{RED}{rgb}{1,0,0}\definecolor{BLUE}{rgb}{0,0,1} 
\RequirePackage[normalem]{ulem} 
\RequirePackage{color}\definecolor{RED}{rgb}{1,0,0}\definecolor{BLUE}{rgb}{0,0,1} 

\DeclareGraphicsExtensions{.eps,.pdf,.jpeg,.jpg,.png}

\newcommand{\bea}{\begin{eqnarray}}
\newcommand{\eea}{\end{eqnarray}}
\newcommand{\be}{\begin{equation}}
\newcommand{\ee}{\end{equation}}

\newcommand*{\muup}           {\ifmmode\mathrm{\mu}\else$\mathrm{\mu}$\fi}
\newcommand*{\nuup}           {\ifmmode\mathrm{\nu}\else$\mathrm{\nu}$\fi}
\newcommand*{\piup}           {\ifmmode\mathrm{\pi}\else$\mathrm{\pi}$\fi}
\newcommand*{\muonp}          {\ifmmode\mathrm{\muup^+}\else$\mathrm{\muup^+}$\fi}
\newcommand*{\muonn}
{\ifmmode\mathrm{\muup^-}\else$\mathrm{\muup^-}$\fi}
\newcommand*{\muon}           {\ifmmode\mathrm{\muup}\else$\mathrm{\muup}$\fi}

\newcommand*{\photon}         {\ifmmode{\gamma}\else${\gamma}$\fi}
\newcommand*{\positron}       {\ifmmode{\mathrm{e}^+}\else${\mathrm{e}^+}$\fi}
\newcommand*{\electron}       {\ifmmode{\mathrm{e}}\else${\mathrm{e}}$\fi}

\newcommand{\meg}{\ifmmode{\muup \to e \gamma}\else$\mathrm{\muup \to e \gamma}$\fi}
\newcommand{\megp}{\ifmmode{\muup^+ \to \mathrm{e}^+ \gamma}\else$\mathrm{\muup^+ \to e^+ \gamma}$\fi}
\newcommand{\michel}{\ifmmode{\muup^+ \to e^+ \nuup\bar{\nuup}}\else$\mathrm{\muup^+ \to e^+ \nuup\bar{\nuup}}$\fi}
\newcommand{\radiative}{\ifmmode{\muup^+ \to \mathrm{e}^+\nuup\bar{\nuup}\gamma} \else$\mathrm{\muup^+ \to e^+ \nuup\bar{\nuup}\gamma}$\fi}

\newcommand*{\mathtentative}{}
\def\mathtentative#1#{\mathcoloraux{#1}}
\newcommand*{\mathcoloraux}[3]{%
  \protect\leavevmode
  \begingroup
    \color#1{#2}#3%
  \endgroup
}

\begin{document}


\title{Constraints on Lorentz invariance from the event KM3-230213A
}

\author{P.W.~Cattaneo }
\affiliation{INFN Pavia, Via Bassi 6, Pavia 27100, Italy}

\date{\today}

\begin{abstract}
Lorentz invariance is the cornerstone of relativity theory. Its implications
have been verified experimentally with a variety of approaches.
The detection of a muon at extremely high energy detected by the ARCA 
detector in the Mediterranean sea, the most energetic particle directly measured 
up to date, allows to put additional constraints on Lorentz non-invariant 
theories. The prediction of some of those theories is that the lifetimes of 
particles in the laboratory frame 'decrease' rather than 'increase' with increasing 
$\gamma$.
In this frame the sheer fact that the muon traversed the whole ARCA detector 
puts a lower limit on the muon lifetime in the laboratory frame, that implies
upper limits on Lorentz violating parameters. 
\end{abstract}

\maketitle 

\section{Introduction}

The paper \cite{Coleman:1998ti} develops a model with small 
violation of the Lorentz invariance which does not respect 
flavour conservation allowing the Maximum Attainable Velocity (MAV) 
for each massive particle to be different from the speed of light 
and different between different particles.
For each massive particle 
and separately for the two helicity states a parameter $\epsilon$ 
defines the relative difference between the MAV and the 
speed of light.\\
A consequence of this model is that the branching 
ratio of the decay \meg\ depends on the muon energy as 
measured in a privileged frame such as the one in which 
the 2.7 K universal microwave background has no dipole anisotropy.
The branching ratio for this decay channel is suppressed 
in the Standard Model to \num{\sim e-54} and its 
experimental upper limit measured at rest at \SI{90}{\percent} C.L. is 
${\cal B} (\megp) < \num{3.1e-13}$ \cite{MEGII:2023ltw}.

For a muon moving with a Lorentz factor $\gamma$ 
the situation changes. 
According to Eq.3.8 and 3.9 in \cite{Coleman:1998ti} the inverse of the total decay length can be written as

\begin{equation}
\frac{1}{l_d(\gamma)}= \Gamma/c = \Gamma_\mathrm{w}/c + \Gamma_{r}/c = 
\frac{1}{\tau_0c\gamma} + \frac{b_{L,R}\gamma^3}{\tau_0c} \label{eq:gammac}
\end{equation}

where $\tau_0$ is the muon rest lifetime in the Standard Model and
$b_{L,R}$ is a parameter describing the violation of the Lorentz 
invariance and flavour conservation 
\footnote[1]{L refers to left handed $\mu^-$ or right handed $\mu^+$, R the opposite. For muon at 
rest $b_{L,R}=b$.}.

For ultrarelativistic muons, it can be written as
\begin{eqnarray}
b_L = \frac{\alpha m_\mu \tau_0}{30}(68\epsilon_R^2 + \epsilon_L^2), \\
b_R = \frac{\alpha m_\mu \tau_0}{30}(68\epsilon_L^2 + \epsilon_R^2).
\end{eqnarray}
where $m_\mu$ is the muon rest mass, $\alpha$ the fine structure constant and 
\begin{equation}
\epsilon^2_{L,R} = |\delta c_{L,R}|^2,
\label{eq:epsilon}
\end{equation}
from Eq.3.4 in \cite{Coleman:1998ti}, where $\delta c_{L,R}$ is the 
helicity dependent relative deviation of muon MAV from 
the speed of light.

From Eq.3.6 in \cite{Coleman:1998ti} the \meg\ 
branching ratio at rest is expected to be
\begin{equation}
b = \frac{\alpha m_\mu \tau_0}{4}(\epsilon_R^2 + \epsilon_L^2),
\label{eq:brest}
\end{equation}

from this equation we can put a limit on $b$
from the above mentioned limit on ${\cal B} (\megp)$
\[
b \approx \num{
6.4 e14} (\epsilon_R^2 + \epsilon_L^2) < \num{3.1e-13}
\]
that implies
\[
(\epsilon_R^2 + \epsilon_L^2) < \num{4.8e-28}.
\]

But the most striking feature of Eq.\ref{eq:gammac}
is that at high $\gamma$ the decay channel 
\meg\ dominates and the muon lifetime decreases quickly with increasing energy.
Therefore this limit is largely overcome in \cite{Cowsik:1998hp} from an analysis 
of cosmic rays induced horizontal showers that obtains the very strong
result of $b_R < \num{e-25}$, that translate in $(\epsilon_R^2 + 68\epsilon_L^2) < \num{e-39}$. 
This result is nevertheless fraught of assumptions on cosmic rays behaviour that seems not appropriate for 
an expected path length shorter than a few kilometers. The limit refers
to $b_R$ because the \muonn\ (\muonp) originating from $\pi$ and $K$ decays are mostly right (left) handed.

\section{The event KM3-230213A from KM3Net}

In \cite{KM3NeT:2025npi} the experiment KM3Net \cite{KM3Net:2016zxf} reports the detection of an extremely high energy muon with energy 
$E_{\mu} = 120 ~ { }^{+ 110}_{-60} \textrm{PeV}$ by the ARCA detector in the 
Mediterranean sea. The muon traverses the full detector for a length of approximately 
$l_\mu \SI{\sim 500}{\meter}$ implying that at about 90 \%\ confidence level the 
decay length is larger than $(l_\mu/2.3) \sim \SI{220}{\meter}$\footnote{The probability density for decay length is $\frac{1}{l_0} \exp (-l/l_0)$ ($l_0$ is the average value.).
For a single measurement $l$ the probability that the average decay length is shorter than $l_0$ is
$\exp (-l/l_0)$. Therefore $l_0>l/2.3$ at 90\%\ C.L. 
because $\exp(-2.3) = 0.10$.}, the most likely value 
for the muon Lorentz factor is
$\gamma = \SI{120}{PeV}/(m_\mu c^2) \sim \num{1.15e9}$ 
and the value 1-$\sigma$ lower is 
$\gamma_\sigma = \SI{60}{PeV}/(m_\mu c^2) \sim \num{0.57e9}$.
From Eq.\ref{eq:gammac} a conservative estimation at 90\%\ C.L. of the limit on $b_L$ is
\[
b_L < \gamma_\sigma^{-3} l_d(1)/(l_\mu/2.3) \sim \num{1.6e-26},
\]
where $l_d(1)\approx \tau_0c$.\\
This limit refers to $b_L$ because a \muonn\ (\muonp) originating from a charged current neutrino interaction is left (right) handed and implies
\[
(68\epsilon_R^2 + \epsilon_L^2) < \num{1.6e-40}.
\]
This limit is substantially better than the one reported in \cite{Cowsik:1998hp} even if it is based on a single event, because of 
the exceptionally high $\gamma$ of the detected muon.

Assuming $\epsilon_R=\epsilon_L$ and maximal velocity mixing between muon and electron, from Eq.\ref{eq:epsilon}
the limit on the difference in MAV between the two leptons is
\[
|c_\mu-c_e| < \num{1.5e-21}
\]

\bibliographystyle{my}
\bibliography{RelativityK3Net}
\end{document}